\documentclass{aa}

\usepackage{graphicx}
\usepackage{txfonts}
\usepackage{longtable}    
\usepackage[colorlinks=true,urlcolor=blue,citecolor=blue]{hyperref}
\usepackage{hyperref}
\newcommand{\hi}{{\rm H{\textsc{i}}\,}}
\newcommand{\hii}{{\rm H{\textsc{ii}}\,}}

\newcommand{\ci}{{\rm C{\textsc{i}}\,}}
\newcommand{\cii}{{\rm C{\textsc{ii}}\,}}
\newcommand{\ciii}{{\rm C{\textsc{iii}}\,}}
\newcommand{\civ}{{\rm C{\textsc{iv}}\,}}
\newcommand{\cv}{{\rm C{\textsc{v}}\,}}
\newcommand{\cvi}{{\rm C{\textsc{vi}}\,}}
\newcommand{\cvii}{{\rm C{\textsc{vii}}\,}}
\newcommand{\oi}{{\rm O{\textsc{i}}\,}}
\newcommand{\oii}{{\rm O{\textsc{ii}}\,}}

\newcommand{\ovi}{{\rm O{\textsc{vi}}\,}}

\newcommand{\nitv}{{\rm N{\textsc{v}}\,}}

\newcommand{\sili}{{\rm Si{\textsc{i}}\,}}
\newcommand{\silii}{{\rm Si{\textsc{ii}}\,}}

\newcommand{\suli}{{\rm S{\textsc{i}}\,}}
\newcommand{\sulii}{{\rm S{\textsc{ii}}\,}}

\newcommand{\fei}{{\rm Fe{\textsc{i}}\,}}
\newcommand{\feii}{{\rm Fe{\textsc{ii}}\,}}

\begin{document}

\title{Relativistic electron impact ionization cross sections of carbon ions and application to an optically thin plasma}
   \author{Miguel A. de Avillez\inst{1,2} \and Mauro Guerra\inst{3} \and Jos\'e Paulo Santos\inst{3} \and Dieter Breitschwerdt\inst{2}}
   \institute{Department of Mathematics, University of \'Evora, R. Rom\~ao Ramalho 59, 7000 \'Evora, Portugal\\
   \email{mavillez@galaxy.lca.uevora.pt}
   \and
   Zentrum f\"ur Astronomie und Astrophysik, Technische Universit\"at Berlin, Hardenbergstrasse 36, D-10623 Berlin, 
   Germany
   \and
   Laboratory of Instrumentation, Biomedical Engineering and Radiation Physics (LIBPhys-UNL), Department of 
   Physics, Faculty of Sciences and Technology, New University of Lisbon, 2829-516 Caparica, Portugal
    }

    \date{Received February 22, 2019; accepted MM DD, YYYY}

   \titlerunning{Relativistic electron-impact ionization cross sections}
   \authorrunning{de Avillez et al.}


\abstract
{Ionization through electron impact is a fundamental process associated with the evolution of the ionic structure and 
emissivity of astrophysical plasmas. Over several decades substantial efforts have been made to measure and calculate the 
ionization cross sections of ionization through electron impact of different ions shell by shell, in particular, of 
carbon ions. Spectral emission codes use electron-impact ionization cross sections and/or rates taken from different 
experimental and theoretical sources. The theoretical cross sections are determined numerically and include a diversity of 
quantum mechanical methods. The electron-impact ionization database therefore is not uniform in the methods, which makes it 
hard to determine the reason for the deviations with regard to experimental data. In many cases only total ionization rates for 
Maxwell-Boltzmann plasmas are available, which makes calculating inner-shell ionization in collisional-radiative models using 
thermal and nonthermal electron distribution functions difficult. A solution of this problem is the capability of generating the 
cross sections with an analytical method using the minimum number of atomic parameters. In this way, uniformity in the 
database is guaranteed, and thus deviations from experiments are easily identified and traced to the root of the method.
}
{The modified relativistic binary encounter Bethe (MRBEB) method is such a simple analytical scheme based on one atomic 
parameter that allows determining electron-impact ionization cross sections. This work aims the determination of K- and 
L-shell cross sections of the carbon atom and ions using the MRBEB method and show their quality by: (i) comparing them with 
those obtained with the general ionization processes in the presence of electrons and radiation (GIPPER) code and 
the flexible atomic code (FAC), and (ii) determining their effects on the ionic structure and cooling of an optically thin plasma.
}
{The MRBEB method was used to calculate the inner-shells cross sections, while the plasma calculations were carried out 
with the collisional+photo ionization plasma emission software (CPIPES). The mathematical methods used in this work comprise 
a modified version of the double-exponential over a semi-finite interval method for numerical integrations, Gauss-elimination 
method with scaled partial pivoting for the solution of systems of linear equations, and an iterative least-squares method to 
determine the fits of ionization cross sections. 
}
{The three sets of cross sections show deviations among each other in different energy regions. The 
largest deviations occur near and in the peak maximum. Ion fractions and plasma emissivities of an optically thin 
plasma that evolves under collisional ionization equilibrium, derived using each set of cross sections, show deviations that 
decrease with increase in temperature and ionization degree. In spite of these differences, the calculations using the three sets of cross sections agree overall.
}       
{A simple model like the MRBEB is capable of providing cross sections similar to those calculated with more 
sophisticated quantum mechanical methods in the GIPPER and FAC codes.
}
\maketitle
\section{Introduction}

Carbon is the fourth most abundant element in nature after H, He, and O. It has a solar abundance of 
$\log[$C/H$]=-3.57$ \citep{agss2009} and thus contributes much to the emissivity of a plasma. Jointly with He, carbon contributes to the second peak maximum observed in the radiative cooling function of a 
plasma that evolves under collisional ionization equilibrium \citep[see, e.g.,][]{shapiro1976,boeringher1989, 
schmutzler1993, sutherland1993, gnat2007, avillez2010}. In addition, \ci and \cii ions (through fine-structure and 
metastable transitions) are the main contributors to radiative cooling at temperatures below $10^{3}$ K. The 
other contributors are \oi, \sili, \silii, \suli, \fei, and \feii through fine-structure lines and \oi, \oii, \sili, \silii, \sili, 
\sulii, \fei, and \feii through metastable transitions \citep[see, e.g.,][]{dalgarno1972,wolfire1995, wolfire2003}. 
Furthermore, carbon plays an important role in the chemistry of molecular clouds \citep[see, 
e.g.,][]{larson1981,pavlovski2002,glover2007} as well as in shocks \citep[see, e.g.,][]{hollenbach1989}.

The cross sections of the electron-impact ionization of carbon ions have been extensively studied by means of 
theoretical calculations and experimental measurements. Since the seminal work of \citet{bethe1930}, who 
derived the correct form of the cross section at high electron energies, much effort has been invested to calculate 
the cross sections using  empirical and semi-empirical methods or more sophisticated quantum mechanical 
numerical calculations. Compilations comprising theoretical and experimental ionization cross sections for the 
carbon atom and ions, including the best fits to these, that have been popular in the Atomic Physics 
and Astrophysics communities, were published over the years by \citet{lotz1967,lotz1968}, 
\citet[][hereafter BL83]{bell1983}, \citet[][AR85]{ar1985}, \citet[][L88]{lennon1988}, \citet[][SK06]{suno2006}, 
\citet[][M07]{mattioli2007}, \citet[][D2007]{dere2007}, to name only a few. 

The recommended total (BL83, L88, SK06 and M07) and partial (AR85) cross sections are parametrized using the 
\citet{younger1981} formula:
\begin{equation}
\sigma(E)=\frac{10^{-13}}{I^{2}_{_{Z,z}} u}\left[\sum_{i=1}^{n_{max}}A_{i}\left(1-\frac{1}{u}\right)^{i}+B\ln(u)+C\frac{\log   
u}{u}\right]\mbox{~cm$^{2}$}
,\end{equation}
where $u=E/I_{_{Z,z}}$, $I_{_{Z,z}}$ is the ionization potential (in eV) of the ion with atomic number $Z$ and ionic 
state $z$, $E$ is the incident electron kinetic energy (in eV); $n_{max}=2$ in \citet{younger1981} and AR85 and 5 
in BL83, SK06 and M07. The coefficient $B$ (the Bethe constant) is determined from the photoionization cross 
section, $\sigma_{ph}$, through 
\begin{equation}
B=4I_{_{Z,z}}\int_{I_{_{Z,z}}}^{\infty} \frac{1}{\epsilon}\frac{df}{d\epsilon}d\epsilon=\frac{I_{_{Z,z}}}{\pi \alpha} 
\int_{I_{_{Z,z}}}^{\infty}\frac{\sigma_{ph}}{\epsilon}d\epsilon,
\end{equation}
where $df/d\epsilon$ is the optical differential oscillator strength. At high energies, the cross section 
tends to the Bethe limit \citep[see, e.g.,][]{younger1981,pradhan2011},
\begin{equation}
\lim_{u\to \infty} \sigma(E)=\frac{1}{u}B\ln(u).
\end{equation}
When $B$ is known, the coefficients $A_{i}$ and $C$ are determined directly from a least-squares fitting 
procedure. Hence, the fit to the cross section has the correct asymptote at high energies. In AR85, the 
ionization potential $I_{_{Z,z}}$ and the coefficients $A_{i}$, $B$, and $C$ refer to the subshell $j$ of the initial 
ion. The total direct ionization cross section is then obtained by summing over all the subshells.

\citet{bell1983} and \citet{lennon1988} adopted the experimental data of \citet{brook1978} for \ci and 
extrapolated the data beyond 1 keV using a fitting equation to the Born approximation. The recommended curve for 
the \cii ionization cross section follows the cross-beam measurements of \citet{aitken1971} and at high 
energies  the Coulomb-Born calculations of \citet{moores1972}. For \ciii  and \civ ionization, Bell and collaborators 
adopted the Coulomb-Born calculations of \citet{jakubowicz1981}. In the case of \ciii ionization, a small 
contribution from inner-shell ionization was included. The cross section for ionization of \cv was obtained by 
scaling  the ionization cross section of {\rm B{\textsc{iv}}\,} along the isosequence. For \cvi the calculations by 
\citet{younger1980a} have been adopted.  

\citet{ar1985} considered the ionization cross sections of the different subshells. The \cvi cross-section (shell 
$1s$) parameters were determined from the distorted wave-exchange approximation calculations of 
\citet{younger1981}. For \cv (shell $1s^2$) a fit to the measured cross section available at the Electron-Impact 
Ionization Data of Multicharged Ions database (EIIDMI) 
\citep{crandall1979report}\footnote{\href{http://www-cfadc.phy.ornl.gov/xbeam}{http://www-cfadc.phy.ornl.gov/xbeam}}
 was carried out. The theoretical values of \citet{younger1981} for the direct ionization from shell $1s^2$ of the Li-like ion \civ 
 were adopted, while for the $2s$ shell ionization, the \citet{crandall1979} data were kept. Excitation-autoiozation contributions 
 to the \civ ion were taken into account; the derived formulae are presented in Appendix A of AR85. For \ciii (shells $1s^2$ and 
 $2s^2$), AR85 adopted the calculations by \citet{younger1981} 
along the sequence. The fitting parameters for the \cii and \ci ionization of shell $2s^2$ were deduced by 
extrapolation from higher-Z elements. The fit to the \cii $2p$ shell ionization cross section is based on the 
measurements of \citet{aitken1971}, and those of \ci were derived from the measurements by \citet{brook1978}. 
The parameter B was derived from the photoionization cross section of \citet{reilman1979}.

\citet{suno2006} adopted for \cvi and \cv the cross sections calculated using the distorted-wave method with 
exchange \citep{pattard1999} and the distorted-wave Born method \citep{fang1995}, respectively. For \civ, the 
experimental data of \citet{knopp2001} were selected. Because of excitation-autoionization,
the cross section has two peaks. For \ciii, the experimental data of \citet{woodruff1978} were chosen, and for \cii and \ci, the 
experimental data of \citet{yamada1989}  and \citet{brook1978} were used, respectively. 

\citet{mattioli2007} adopted cross sections from theoretical and experimental data: the \cvi and \cv cross sections were 
taken from \citet{ar1985}. For \civ, the \citet{crandall1979} data were complemented with those available at the 
EIIDMI database. For \ciii, the data of Falk et al. (1983) for low metastable contributions and that of \citet{loch2005} 
were adopted. Similarly to SK06, the cross sections of \cii and \ci were taken from \citet{yamada1989} and 
\citet{brook1978}, respectively. 

The \citet{dere2007} compilation combines experimental cross sections with those obtained with the flexible atomic 
code (FAC)\footnote{\href{http://www-amdis.iaea.org/FAC}{http://www-amdis.iaea.org/FAC}} 
\citep{gu2002,gu2008}. For \cvi, D2007 favored the parametric fit of \citet{fontes1999} to their relativistic distorted -wave approximation cross-section calculations for ionization from the $1s$ shell; the \cv cross section is 
determined with the FAC. For \civ and \ciii, direct ionization and excitation-autoionization cross sections calculated 
with the FAC were adopted; for \ciii, the $1s$ and $2s$ subshells were taken into account for the direct ionization cross 
sections, while for EA the $1s2l^{3}$ and $1s2l^{2}3l^{\prime}$ transitions were considered. The \citet{yamada1989} 
and \citet{brook1978} cross-section measurements were adopted for \cii and \ci, respectively. The total cross 
sections, available in version 8.0.7 of the CHIANTI atomic 
database\footnote{\href{http://www.chiantidatabase.org}{http://www.chiantidatabase.org}} \citep{delzanna2015}, 
are provided as spline nodes for a scaled energy $U$ and cross section $\Sigma_{_{Z,z}}$ given by
\begin{equation}
U=1-\frac{\log f}{\log(u-1+f)}
\end{equation} 
and 
\begin{equation}
\Sigma_{_{Z,z}}=\frac{u \sigma_{_{Z,z}}I_{_{Z,z}}}{\log(u)+1},
\end{equation} 
respectively. In these expressions $f=2$ is an adjustable parameter, $\sigma_{_{Z,z}}$ is the unscaled cross 
section, and $u$, and $I_{_{Z,z}}$ have the meanings related in previous paragraphs.

Further calculations using sophisticated quantum mechanical methods have been carried out in the last two 
decades by \citet{bote2009}, \citet{abdel-naby2013}, and \citet{wang2013} for \ci, \citet{ludlow2008}, \citet{ballance2011}, 
\citet{pindzola2012}, and \citet{lecointre2013} for \cii, \citet{fogle2008} for \ciii, \citet{pindzola2012} for \civ, and 
\citet{fontes1999} for \cvi. Some of these authors also reported experimental measurements that were used to 
compare with the theoretical cross sections, for instance, \citet{wang2013} for \ci, \citet{lecointre2013} for \cii, and 
\citet{fogle2008} for \ciii. 

Empirical classical and semiclassical analytical methods have also been adopted over the years to calculate ionization cross sections of different ions. These include the binary-encounter-Bethe 
\citep[BEB][]{kim1994} model\footnote{BEB combines the Mott cross section with the high-incident energy 
behavior of the Bethe cross section.} and its derivatives, for example, the relativistic BEB \citep{kim2000} and the modified 
relativistic BEB \citep[][MRBEB]{guerra2012} models. The advantage of these models is their simple analytical 
expressions and the small number of adopted parameters that depend on the binding energy, on the energy of the 
impacting electron, and on the shielding by inner electrons, for instance. The MRBEB model has been applied to calculate the K-, L- and M-shell ionization cross sections of several atoms, with Z varying between 6 and 83 
\citep{guerra2012}, and for several ionization stages of Ar, Fe, and Kr \citep{guerra2013}, and U \citep{guerra2015}. 
The model provides reliable direct ionization cross sections, and the relative differences to experimental data 
are smaller than 10\% for the inner shells of neutral atoms and 20\% for highly charged ions \citep{guerra2012,guerra2013}. 
For a review of these models, see \citet{llovet2014}, for example. 

We here use the MRBEB method to calculate the K- and L-shell cross sections of the carbon ions and 
convolve them with the Maxwell-Boltzmann electron distribution function in order to obtain the corresponding 
ionization rates. A further application is made to the evolution of an optically thin plasma in order to obtain the 
radiative losses due to electron-impact ionization, bremsstrahlung, and line emission. The structure of this paper is 
as follows. Section 2 describes the MRBEB model and the calculation of the K- and L-shell ionization cross 
sections of the carbon atom and ions. Section 3 describes the use of the calculated cross sections in 
determining the ionization structure and radiative losses of an optically thin plasma that evolves under collisional 
ionization equilibrium. Section 5 closes the paper with a discussion and final remarks. Appendix A describes the 
tabulated data.

\begin{figure*}[thbp]
        \centering
        \includegraphics[width=0.42\hsize,angle=0]{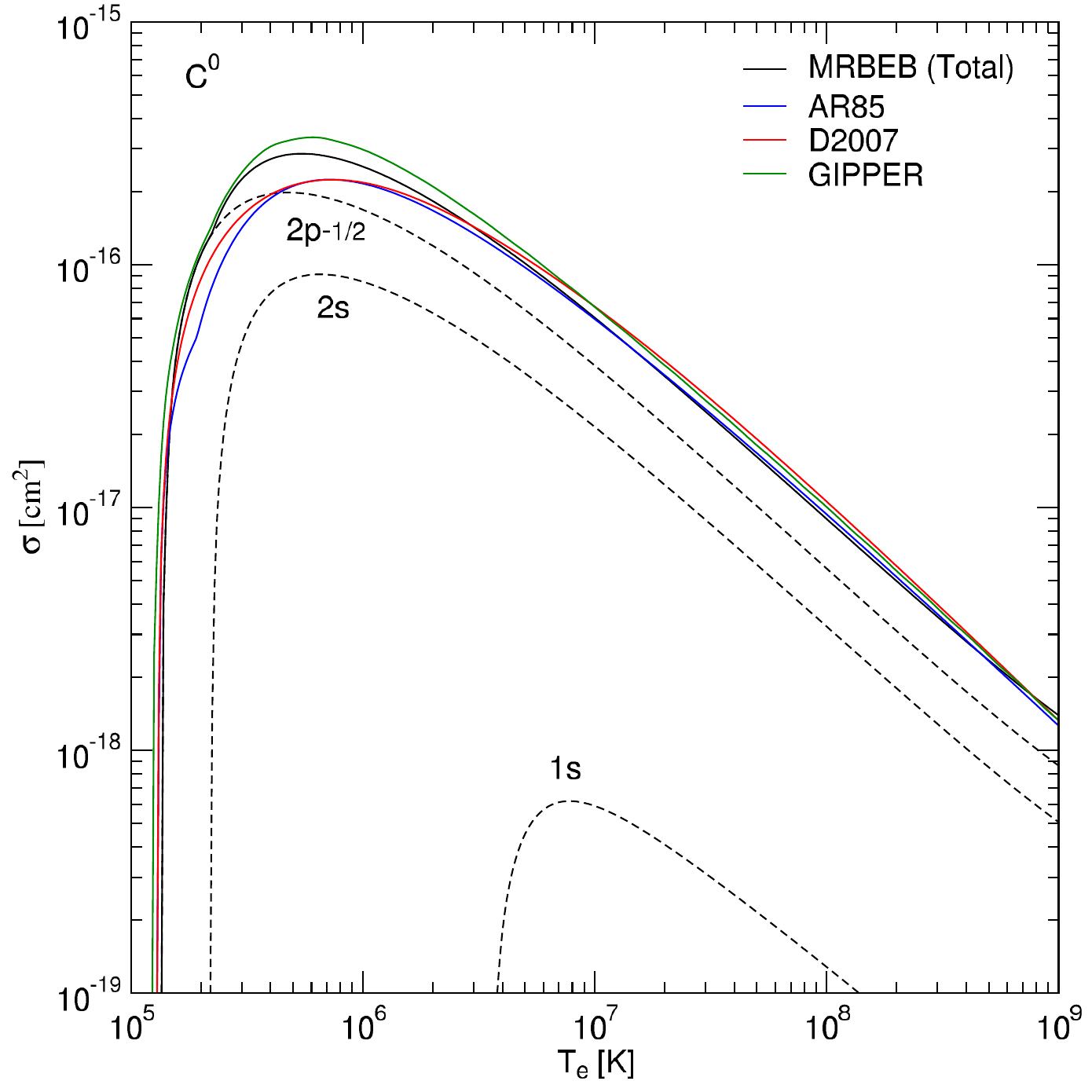}
        \includegraphics[width=0.42\hsize,angle=0]{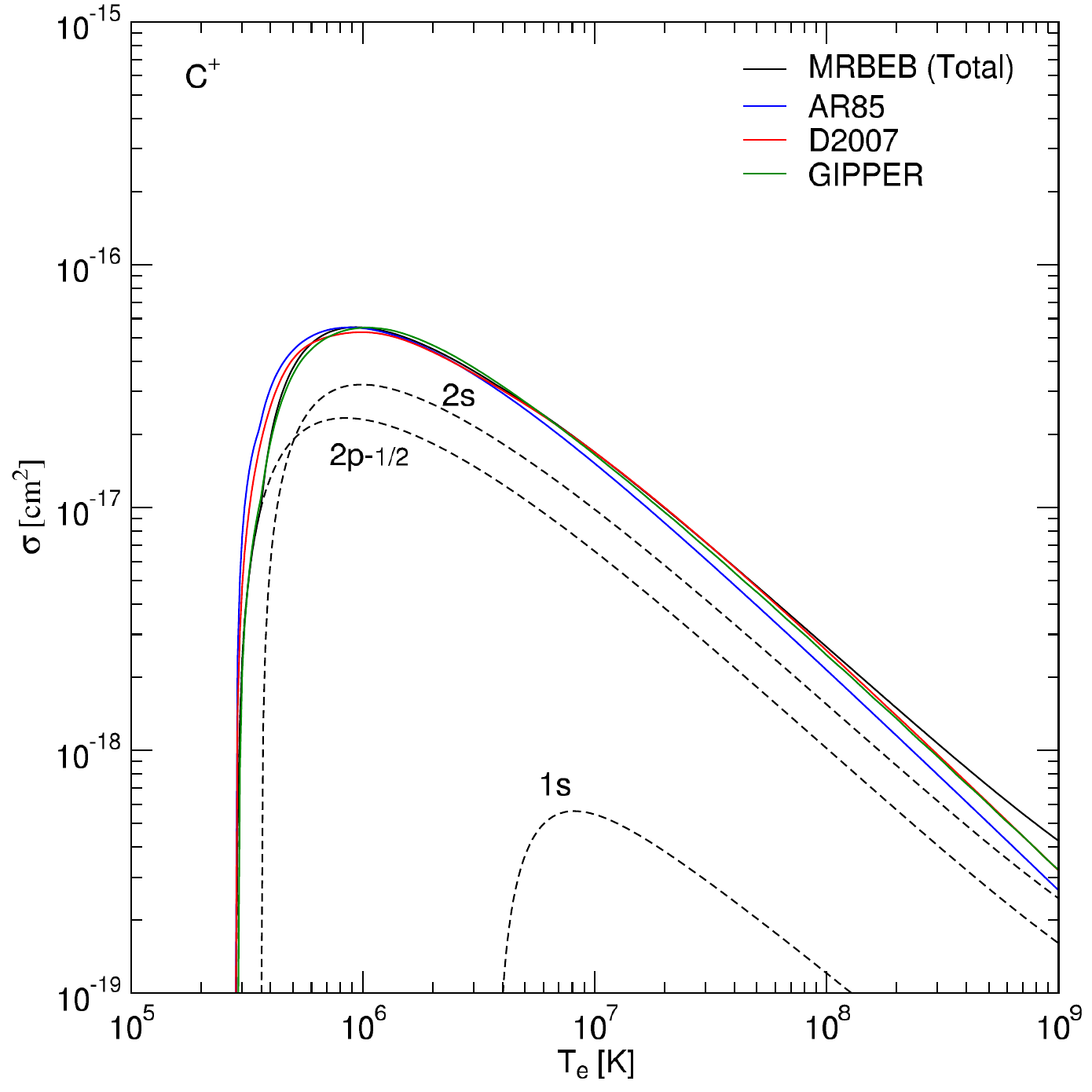}\\
        \includegraphics[width=0.42\hsize,angle=0]{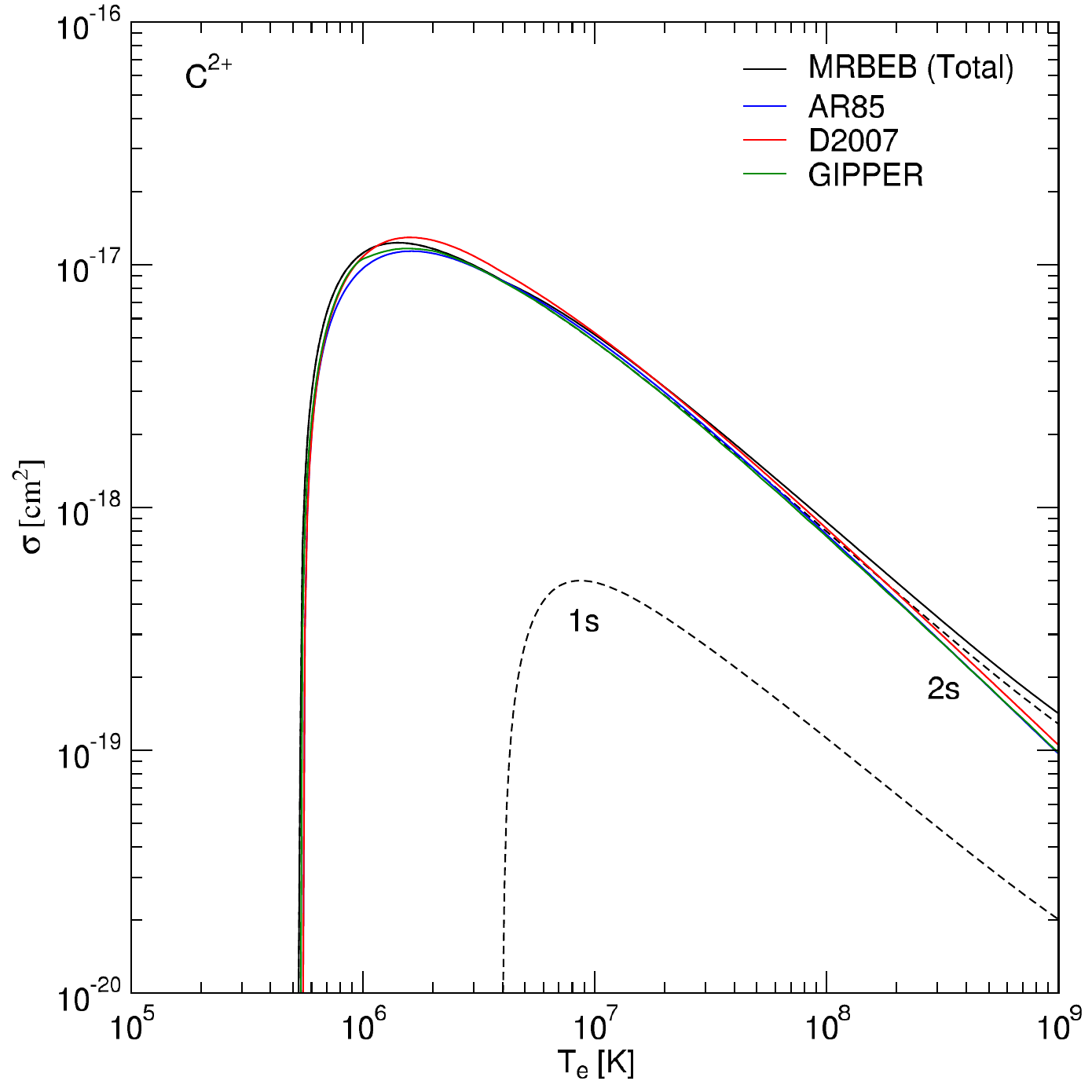}
        \includegraphics[width=0.42\hsize,angle=0]{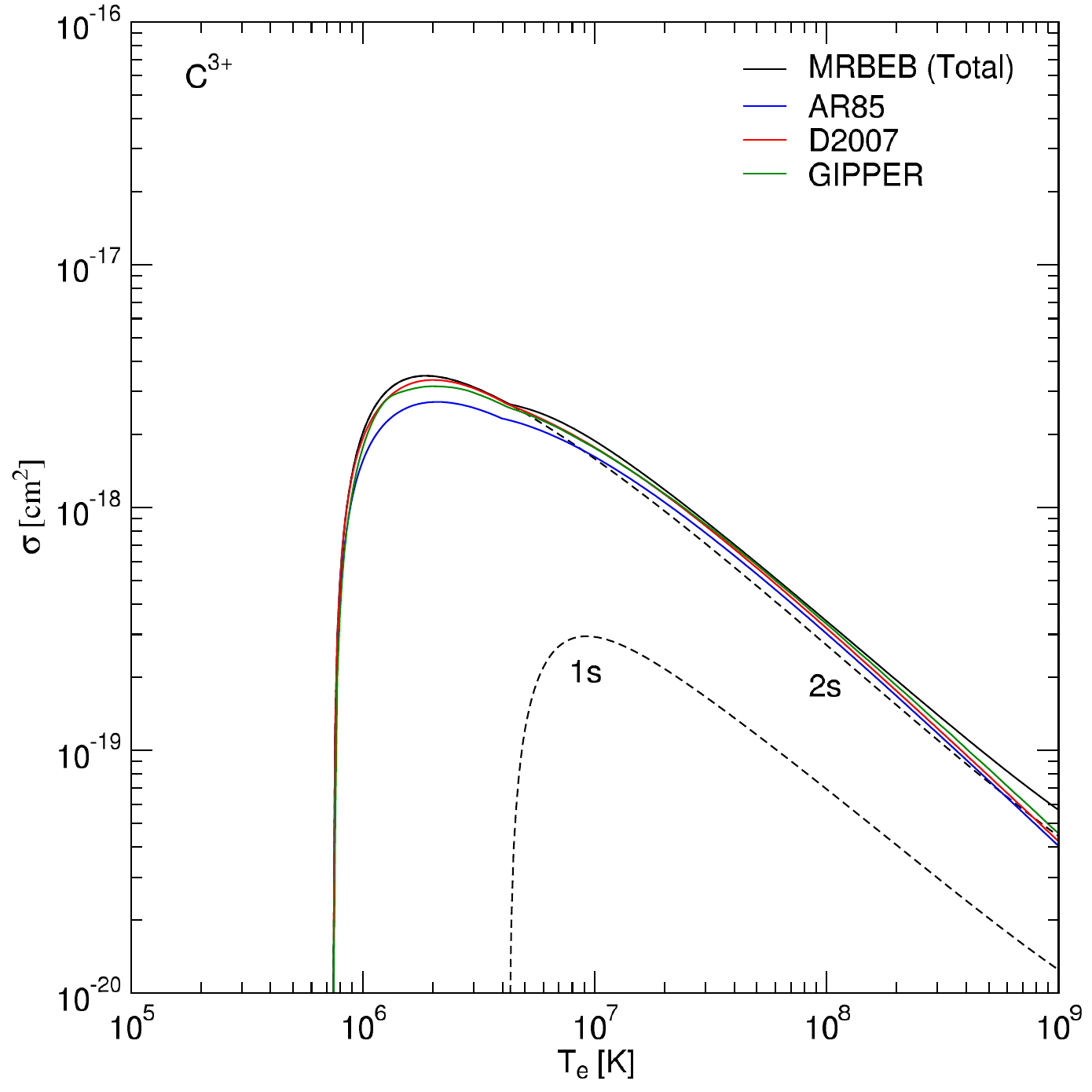}\\
        \includegraphics[width=0.42\hsize,angle=0]{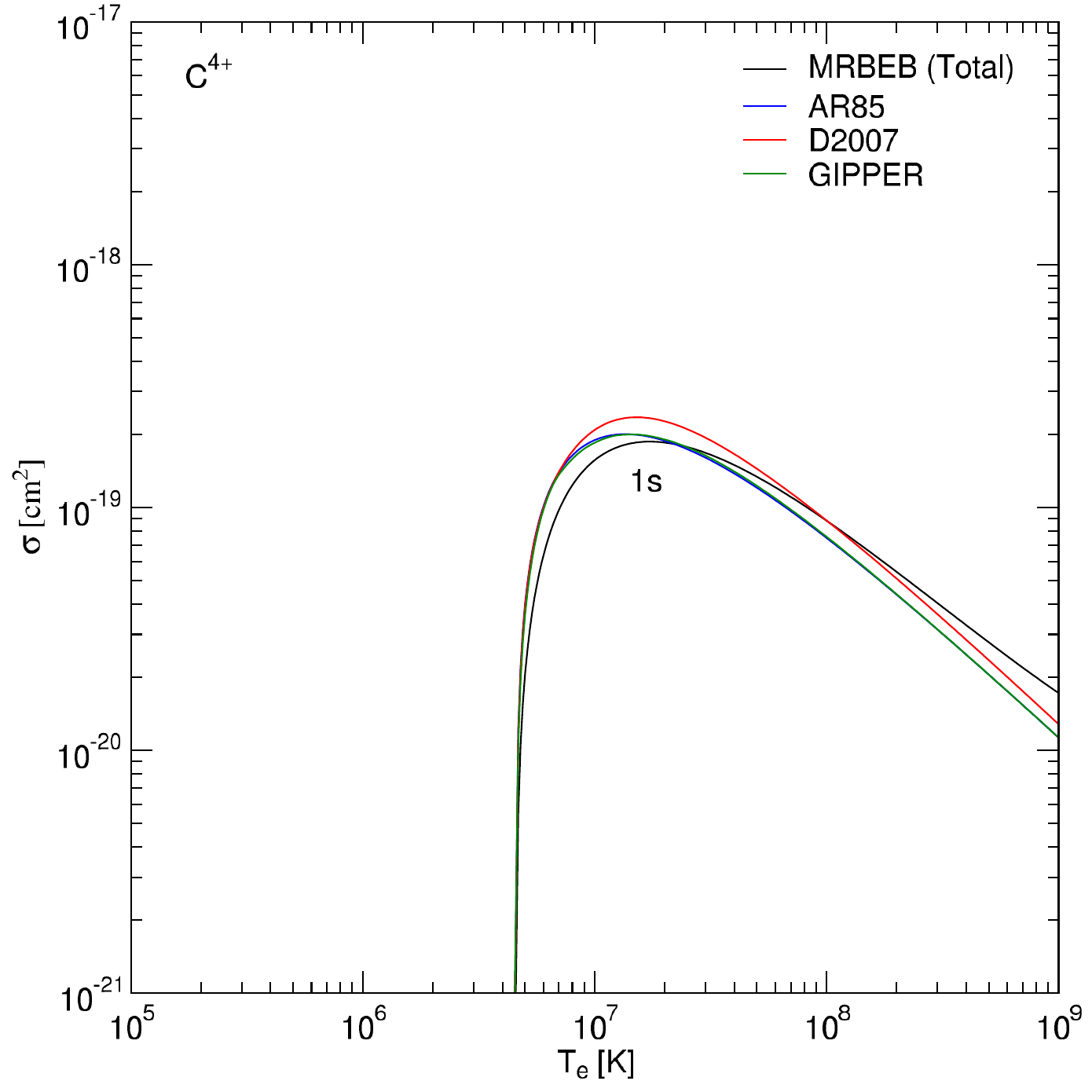}
        \includegraphics[width=0.42\hsize,angle=0]{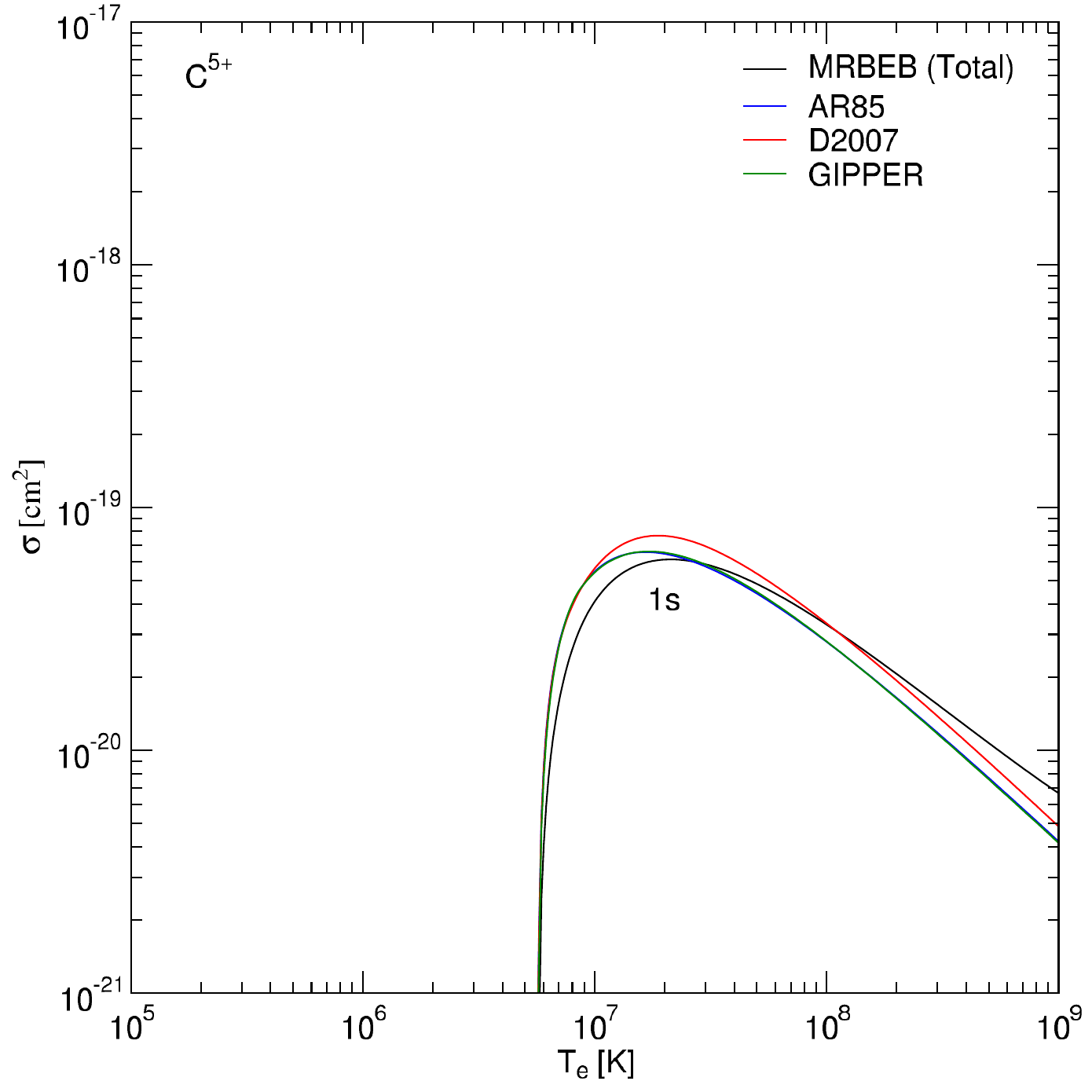}
        \caption{K- and L-shell (dashed black lines) and total (solid black lines) ionization cross sections of the carbon atom and 
        ions calculated with the MRBEB model and total cross sections in AR85 (blue lines) and D2007 (red lines) calculated with the 
        GIPPER code (green lines).\label{mrbeb1}}
\end{figure*}

\section{Modified relativistic binary encounter Bethe model cross sections}

\subsection{The MRBEB model}

The MRBEB model uses an analytical approach containing a single atomic parameter (the binding energy of the electron to be 
ionized) and takes into account the energy of the impacting electron and the shielding of the nuclear charge by the bound 
electrons of the target ion. Therefore, the number of electrons in the inner shells up to the subshell that is ionized acts as a 
screening of the nuclear potential as seen by the primary electron. In contrast to the MRBEB, the other binary encounter Bethe 
and relativistic binary encounter Bethe models require two input parameters (the binding energy and the kinetic energy of the 
bound electron).

The MRBEB cross section, denoted by $\sigma_{nlj,\,LS}$, refers to the ionization of an \textit{nlj} electron in an 
atom or ion in a given initial state LS. The cross section takes into account the relativistic interactions between the 
incident and target electrons during inner-shell ionization of heavy atoms or ions. The determination of the cross 
section requires knowledge of one single parameter: the binding energy of the target electron. The cross 
section is given by \citep{guerra2012}
\begin{equation}
\sigma_{nlj,\,LS} = \frac{4\pi a_{o}^{2}\alpha^{4}N}{\left(\beta_{t}^{2}+\chi \beta_{b}^{2}\right)2 b^{\prime}} \left[A(\beta_{t}, t,
b^{\prime})+B(t, t^{\prime},b^{\prime})\right] \mbox{ cm$^{2}$}
\end{equation}
with
\begin{equation}
A(\beta_{t},t,b^{\prime})=
0.5\left[\ln\left(\frac{\beta_{t}^{2}}{1-\beta_{t}^{2}}\right)-\beta_{t}^{2}-\ln(2b^{\prime})\right]\left(1-\frac{1}{t^{2}}\right)
\end{equation}
and
\begin{equation}
B(t, t^{\prime},b^{\prime})=1 -\frac{1}{t}-\frac{\ln t}{1+t}\frac{1+2t^{\prime}}{(1+t^{\prime}/2)^{2}}+
\frac{b^{\prime\, 2}}{(1+t^{\prime}/2)^{2}}\frac{t-1}{2}
\end{equation}
with
\begin{eqnarray}
t=\frac{E}{B}, & \beta_{t}^{2}=1-\frac{1}{(1+t^{\prime})^{2}}, & b^{\prime}=\frac{B}{m_{e} c^{2}}, \nonumber \\
t^{\prime}=\frac{E}{m_{e} c^{2}},  & \beta_{b}^{2}=1-\frac{1}{(1+b^{\prime})^{2}}. &  \nonumber
\end{eqnarray}
In these equations, $E$ denotes the kinetic energy of the impacting electron, $B$ is the binding energy of the 
target electron, and $c$ stands for the speed of light, while $m_{e}$ is the electron mass; $a_{o}$ 
and $\alpha$ are the Bohr radius and the fine-structure constant, respectively. The scaling $\displaystyle 
\frac{1}{\beta^{2}+\chi \beta_{b}^{2}}$ includes the effects of the shielding of the nucleus by the target-bound 
electron through the parameter $\chi$ , which is therefore related to the shielding coefficient $C_{nlj}(Z)$ 
and the binding energy $B$ of the target electron through the relation $\chi=2\,\mbox{Ry}\, C_{nlj}(Z)/B$ (Ry is 
the Rydberg energy). The shielding of the nucleus is described by
\begin{equation}
C_{nlj}(Z)=0.3 \frac{Z^{2}_{eff,\,nlj}}{2n^{2}}+0.7 \frac{Z^{2}_{eff,\,n'l'j'}}{2n^{\prime 2}}
,\end{equation}
where $n^{\prime}l^{\prime}j^{\prime}$ stands for the next subshell after subshell $nlj$, ordered in energy, and 
$Z_{eff}$ is the screening effect \citep[][]{guerra2017}.

\subsection{Cross-section calculations}

Using the MRBEB model, we calculated the K- and L-shell ionization cross sections for carbon and its ions. The 
binding energies of each target electron were calculated using the multi-configuration Dirac-Fock (MCDF) 
theoretical framework with the multi configuration Dirac-Fock and general matrix element (MCDFGME) code 
\citep{indelicato1990}, which evaluates level energies, including 
correlation and first- and second-order quantum electrodynamics corrections. Table~\ref{binding_energies} 
displays the binding energies obtained with the MCDF method without electronic correlation beyond the 
intermediate coupling using the \citet{rodrigues2004} ground-state configurations for the different ions. First-order 
retardation terms of the Breit operator and the Uelhing contribution to the vacuum polarization terms were 
included self-consistently. The Wichmann-Kroll and Kallen-Sabry contributions, as well as higher-order Breit 
retardation terms and other QED effects, such as self-energy, were included as perturbations.
\begin{table}[thbp]
        \centering
        \caption{Binding energies (eV) calculated with the MDFGME code for the ground-state configurations of \ci-\cvi.\label{binding_energies}}
        \begin{tabular}{p{0.22cm}p{1.33cm}p{1.3cm}p{1.1cm}p{1.1cm}p{1.1cm}}
                \hline
                \hline
                \noalign{\smallskip}
                        Ion & ~~~Config. & \multicolumn{4}{c}{Shells}\\
                        & & ~~~~~~~$1s$ & ~~~~~$2s$ & ~~~$2p_{1/2}$ & ~~~$2p_{3/2}$ \\
                        \hline
                        \noalign{\smallskip}
                        C$^{0}$ & $1s^{2}2s^{2}2p^{2}$  & 308.24671 & 19.21208 & 11.78503 & 11.79174 \\
                        C$^{1+}$ & $1s^{2}2s^{2}2p^{1}$ & 323.82728& 31.40751  & 24.61858 &                \\
                    C$^{2+}$ & $1s^{2}2s^{2}$           & 344.30589& 46.11313   &                   &               \\
                    C$^{3+}$ & $1s^{2}2s^{1}$            & 366.88819& 64.37863 &                 &               \\
                    C$^{4+}$ &  $1s^{2}$                     & 391.39542 &                    &                 &               \\
                    C$^{5+}$ & $1s^{1}$                       & 490.01853 &                    &                 &               \\
                   \hline
                \end{tabular}
        \end{table}

\begin{table*}[thbp]
        \centering
        \caption{$a_{1}$- $a_{5}$ fit coefficients for Eq. (\ref{eq10}).\label{table2}}
        \begin{tabular}{cccrccccccc}
                \hline \hline
                Z & Ion & Shell & E$_{nlj}$ (eV)& a$_{1}$ & a$_{2}$ & a$_{3}$ & a$_{4}$ & a$_{5}$ & $\Delta \sigma^{-}$ [\%]& $\Delta 
                \sigma^{+}$ [\%]\\
                \hline
                6 &  0 & 1s          & 308.25 &   9.2135E-2  &   5.2082E-4  &  -1.1545E-1  &   1.1282E-1  &  -4.2654E-1  &  -0.262 &   0.157 \\
                6 &  0 & 2s          &  19.21 &   1.4319E+0  &   9.4133E-3  &  -2.3794E+0  &   3.1997E+0  &  -7.4347E+0  &  -0.446 &   0.915 \\
                6 &  0 & 2p$_{1/2}$ &  11.79 &   2.2848E+0  &   1.6892E-2  &  -4.1305E+0  &   6.4108E+0  &  -1.3503E+1  &  -0.691 &   1.315\\
                6 &  1 & 1s          & 323.83 &   8.7667E-2  &   4.9960E-4  &  -1.0947E-1  &   1.0643E-1  &  -4.0545E-1  &  -0.257 &   0.152 \\
                6 &  1 & 2s          &  31.41 &   8.8915E-1  &   5.2590E-3  &  -1.3856E+0  &   1.6595E+0  &  -4.2079E+0  &  -0.341 &   0.400 \\
                6 &  1 & 2p$_{1/2}$ &  24.62 &   7.9808E-1  &   4.8760E-3  &  -1.2868E+0  &   1.6485E+0  &  -3.9822E+0  &  -0.370 & 0.598\\
                6 &  2 & 1s          & 344.31 &   8.2511E-2  &   4.6848E-4  &  -1.0285E-1  &   1.0100E-1  &  -3.8499E-1  &  -0.264 &   0.156 \\
                6 &  2 & 2s          &  46.11 &   6.0642E-1  &   3.5623E-3  &  -9.4876E-1  &   1.1577E+0  &  -2.9084E+0  &  -0.326 &   0.808 \\
                6 &  3 & 1s          & 366.89 &   7.7408E-2  &   4.4358E-4  &  -9.6027E-2  &   9.4031E-2  &  -3.6139E-1  &  -0.257 &   0.151 \\
                6 &  3 & 2s          &  64.38 &   3.0939E-1  &   1.7414E-3  &  -4.6251E-1  &   5.3552E-1  &  -1.4540E+0  &  -0.266 &   0.372 \\
                6 &  4 & 1s          & 392.40 &   7.1514E-2  &   4.4140E-4  &  -1.0432E-1  &   1.1348E-1  &  -3.2384E-1  &  -0.302 &   0.375 \\
                6 &  5 & 1s          & 490.02 &   4.0675E-2  &   2.4722E-4  &  -5.8541E-2  &   6.5142E-2  &  -1.9087E-1  &  -0.271 &   0.404 \\
                \hline
        \end{tabular}
\end{table*}

\begin{table*}[thbp]
        \caption{$b_{1}$- $b_{5}$ fit coefficients for Eq. (\ref{eq11}).\label{table3}}
        \begin{tabular}{cccrccccccc}
                \hline \hline
                Z & Ion & Shell & E$_{nlj}$ (eV)& b$_{1}$ & b$_{2}$ & b$_{3}$ & b$_{4}$ & b$_{5}$ & $\Delta \sigma^{-}$ [\%]& $\Delta 
                \sigma^{+}$ [\%]\\
                \hline
                6 &  0 & 1s         &  308.25 &   2.3413E-06  &   1.8611E-08  &   2.0518E-05  &  -1.0142E-07  &   1.0674E-08  &  -0.037 &   0.013 \\
                6 &  0 & 2s         &   19.21 &   3.7588E-05  &   1.5327E-07  &   4.3387E-04  &  -1.3627E-06  &   3.3438E-07  &  -0.029 &  0.031 \\
                6 &  0 & 2p$_{1/2}$ &   11.79 &   6.1268E-05  &   2.3858E-07  &   7.3742E-04  &  -2.3593E-06  &   5.7687E-07  &  -0.049 &   0.070 \\
                6 &  1 & 1s         &  323.83 &   2.2285E-06  &   1.8076E-08  &   1.9421E-05  &  -9.8649E-08  &   9.8808E-09  &  -0.037 &   0.013 \\
                6 &  1 & 2s         &   31.41 &   2.2992E-05  &   1.0264E-07  &   2.5401E-04  &  -8.0300E-07  &   1.9680E-07  &  -0.030 &   0.028 \\
                6 &  1 & 2p$_{1/2}$ &   24.62 &   1.4666E-05  &   6.2439E-08  &   1.6563E-04  &  -5.2069E-07  &   1.2779E-07  &  -0.028 & 0.029 \\
                6 &  2 & 1s         &  344.31 &   2.0958E-06  &   1.7286E-08  &   1.8137E-05  &  -9.3739E-08  &   8.5368E-09  &  -0.035 &  0.014 \\
                6 &  2 & 2s         &   46.11 &   1.5661E-05  &   8.7277E-08  &   1.6699E-04  &  -5.6993E-07  &   1.4656E-07  &  -0.014 &   0.020 \\
                6 &  3 & 1s         &  366.89 &   1.9666E-06  &   1.6578E-08  &   1.6896E-05  &  -9.0602E-08  &   7.6196E-09  &  -0.034 &    0.015 \\
                6 &  3 & 2s         &   64.38 &   5.6097E-06  &   3.2241E-08  &   5.7912E-05  &  -1.8881E-07  &   4.7588E-08  &  -0.052 &   0.013 \\
                6 &  4 & 1s         &  392.40 &   1.8395E-06  &   2.7825E-08  &   1.5640E-05  &  -6.9840E-08  &   5.8958E-09  &  -0.008 &  0.007 \\
                6 &  5 & 1s         &  490.02 &   7.3612E-07  &   1.1832E-08  &   6.0987E-06  &  -3.1513E-08  &   1.0332E-09  &  -0.008 &   0.008 \\
                \hline
        \end{tabular}
\end{table*}

Fig.~\ref{mrbeb1} displays the inner shell (dashed black lines) and total (solid black lines) cross sections of the 
carbon ions calculated with the MRBEB model. In addition, the figure also shows the total cross sections of 
\citet[][blue lines]{ar1985}, \citet[][red lines]{dere2007}, and those calculated with the GIPPER code\footnote{GIPPER provides 
electron-impact ionization, photoionization, and autoionization data using the distorted-wave approach.} \citep[][green 
lines]{fontes2015}. In general, the total MRBEB, GIPPER, and D2007 cross sections have a similar distribution for \cii, \ciii, and 
\civ ions. Small deviations occur during the approach to the peak maximum (\cii), in the peak maximum 
(\ciii ans \civ), and after the peak maximum (\ciii and \civ). The total MRBEB cross sections for \cii - \cvi ions 
dominate the others for $T>10^{8}$ K because of the relativistic effects, which are taken into account in the 
calculations.

The MRBEB and GIPPER cross sections have small deviations for \ci, \cii, \ciii, and \civ near and at the peak 
maximum.  The largest deviations between the MRBEB and GIPPER cross sections occur for \cv and \cvi ions 
starting at the approach to the peak maximum. The GIPPER cross sections overlap those discussed in AR85. The 
cross sections in D2007 have the highest values for \cv and \cvi ions at peak maximum and at high energies up to 
$10^{8}$ K when the MRBEB cross section takes over. Clearly, a simple analytical method depending on a single atomic parameter 
gives similar results to those provided by the more sophisticated GIPPER and FAC methods.

\subsection{Fits to the cross sections}
\begin{figure}[thbp]
        \centering
        \includegraphics[width=0.9\hsize,angle=0]{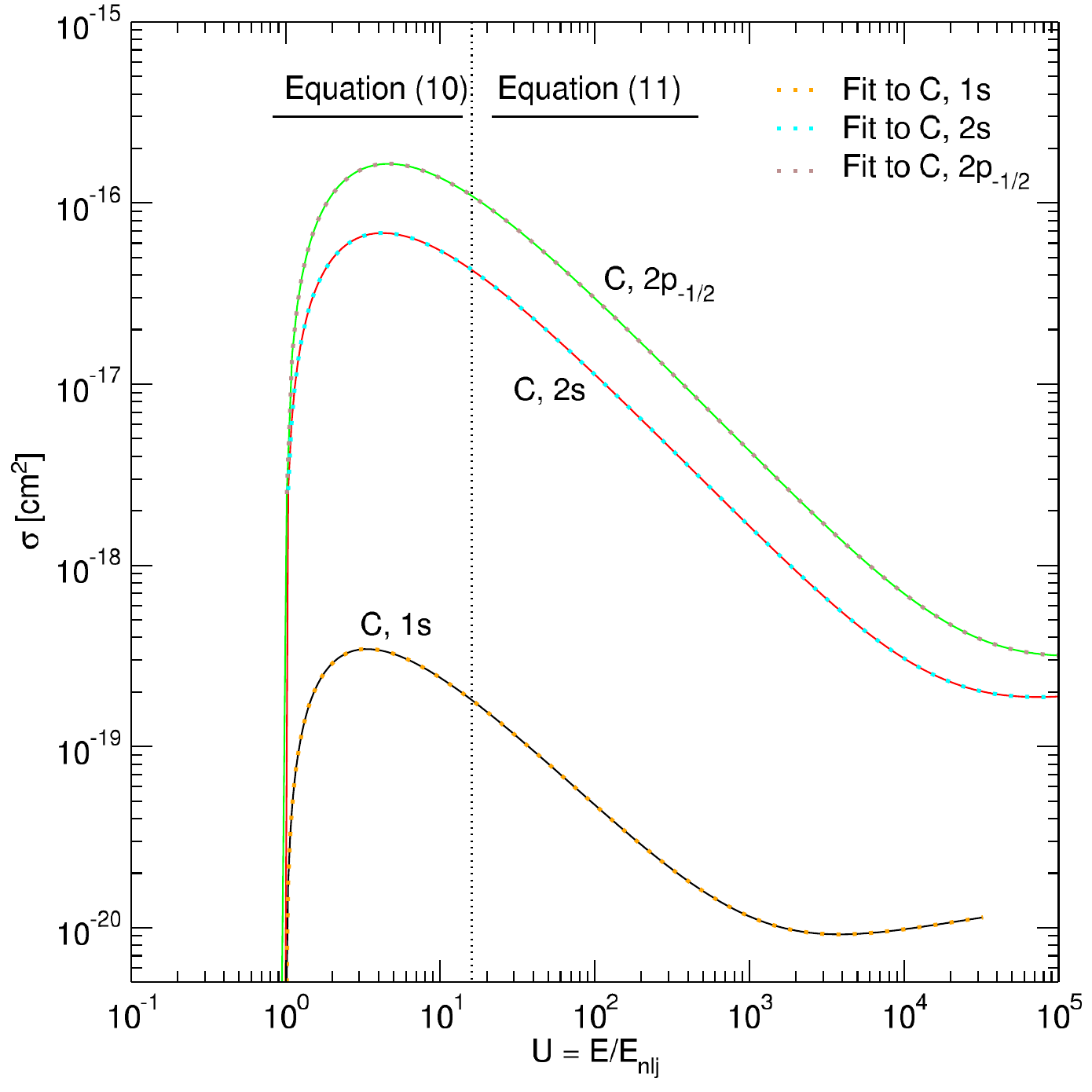}
        \caption{Fit (dotted lines) to the electron-impact ionization cross sections (solid lines) of \ci using Eqs. (\ref{eq10}) and 
        (\ref{eq11}).\label{figfit}}
\end{figure}

The MRBEB cross sections were fit (for \ci see Fig.~\ref{figfit}) with the functional \citep{bote2009}
\begin{equation}
\label{eq10}
\sigma(E)=4\pi a_{o}^{2} 
\frac{U-1}{U^{2}}\left(a_{1}+a_{2}U+\frac{a_{3}}{1+U}+\frac{a_{4}}{(1+U)^{3}}+\frac{a_{5}}{(1+U)^{5}}\right)^{2}\end{equation}
for $U=E/E_{nlj}\leq 16$, while for higher energies a variation of their equation (5) was used,
\begin{equation}
\label{eq11}
\sigma(E)=\frac{U}{U+1.513}\frac{4\pi a_{o}^{2}}{\beta^{2}}\left[ b_{1} \chi 
+b_{2}\frac{\chi}{P}+b_{3}+b_{4}\left(1-\beta^{2}\right)^{1/4}+b_{5}\frac{1}{P}\right],
\end{equation}
with $\chi=2\ln P-\beta^{2}$, and where $\beta=v/c=\sqrt{E(E+2m_{e}c^{2})}/(E+m_{e}c^{2})$ and 
$P=p/(m_{e}c)=\sqrt{E(E+2m_{e}c^{2})}/m_{e}c^{2}$ are the velocity and momentum of the impacting electron, 
respectively, $a_{0}$ is the Bohr radius and the parameters $a_{1}- a_{5}$ and $b_{1}-b_{5}$ (displayed in 
Tables~\ref{table2} and \ref{table3}) are characteristic of each element and electron shell. These parameters were calculated using a least-squares fit. The two right columns in the tables represent the maximum relative differences below 
and above 0, respectively. The absolute relative difference is the largest of the absolute of these two maxima. 
The relative difference is given by
\begin{equation}
\Delta \sigma= \left(1.0-\frac{\sigma_{_{fit}}}{\sigma}\right)100\%.
\end{equation}

The maximum relative differences above and below zero are labeled $\Delta \sigma^{-}$ and $\Delta \sigma^{+}$, 
respectively. The maximum absolute relative differences occur for $U=E/E_{nlj}\leq 16$, that is, with the fitting of 
Eq. (\ref{eq10}). The maximum error is found to be 1.315\% and occurs for the cross section associated with 
shell $2p_{1/2}$ of \ci. This is followed for absolute relative differences of 0.915\% and 0.81\% for the fit of the EII 
cross sections of the $2s$ shell of \ci and \ciii. For the fits associated with energies of $U=E/E_{nlj}> 16,$ the 
maximum absolute relative difference is lower than 0.07\%.

\section{Application to an optically thin plasma} 

We discussed different sets of cross sections for the ionization of the carbon atom and ions and now 
compare their effects on the ionization structure and on the cooling of an optically thin plasma characterized by 
a Maxwell-Boltzmann (MB) electron distribution function.  The determination of the ionic state of the plasma as well 
as its emissivity is important for theoretical models as well as observations of absorption features in the diffuse 
medium where Li-like ions (\civ, \nitv, and \ovi) are important to distinguish between different ionization mechanisms, 
such as turbulent mixing layers \citep{slavin1993}, shock ionization \citep{dopita1996}, conduction interfaces 
\citep{borkowski1990}, and radiative cooling \citep{edgar1986}. These processes in turn are used to study high-velocity clouds \citep[see, e.g.,][]{indebetouw2004} or gas in the Local Bubble \citep[][]{avillez2009}. In addition, the 
emission caused by \cv and \cvi  is important to understand the spectra associated with the processes described 
above, as well as with that of ionizing (\citet{masai1984}) or recombining \citep{avillez2012a} plasmas or in the soft X-ray 
spectra that are observed in supernova remnants. Spectral fitting codes for X-ray emitting space plasmas are often used to 
determine the temperature and ionization state of a plasma. Orbiting X-ray observatories such as Chandra and 
XMM-Newton have chip-based X-ray detectors that can barely resolve individual lines for diffuse low surface brightness plasmas 
like the hot interstellar medium. Therefore it is important to quantify as well as possible the contribution of various ions such as 
carbon to the total spectrum.

For each set of cross sections we followed the evolution of a gas parcel cooling under collisional ionization equilibrium conditions from an initial temperature of $10^{9}$ K where it is completely ionized. The plasma was composed of hydrogen and carbon with solar abundances \citep{agss2009} and a hydrogen particle density, $n_{H}$, of 1 cm$^{-3}$. These calculations are 
referred to as the MRBEB, GIPPER, and D2007 models.

The processes we took into account are electron-impact ionization, radiative recombination including cascades into 
the ground state, dielectronic recombination,  bremsstrahlung, and line emission. No charge-exchange reactions 
were considered. Hence, and because the evolution is in collisional ionization equilibrium, the ionization structure 
of each element (composed of atoms and ions) can be treated independently of the other elements. Thus, the 
evolution of the carbon atom and ions is the same whether the plasma is composed only of carbon or of any set of 
elements including carbon.

\subsection{Thermal model}

The ionization structure and emission properties of the gas parcel were followed using the thermal model 
described in \citet[][; for further details, see de Avillez 2019 in preparation]{avillez2018}. We therefore present a summary of this model in 
this subsection.

The density $n_{Z,z}$ (cm$^{-3}$) of an ion with atomic number $Z$ and charge state $z$ ($z=0,\,...,\,Z$) is 
determined from the populations of the neighboring charge states $z-1$, $z$, and $z+1$ through recombination 
($\alpha_{Z,z}$, which includes radiative and dielectronic recombination) and collisional ionization ($S_{Z,z}$) 
rates from state $z$ to $z-1$ and $z+1$, respectively. Hence, $n_{Z,z}$ is given by the system of equations
\begin{equation}
\label{ionfrac}
S_{Z,z-1}n_{Z,z-1}n_{e}-(S_{Z,z}+ \alpha_{Z,z})n_{Z,z}n_{e}+\alpha_{Z,z+1}n_{Z,z+1}n_{e}=0,
\end{equation}
where $n_{e}$ is the electron density (cm$^{-3}$), which is obtained from
\begin{equation}
\label{electron}
n_{e}=\sum_{z=1}^{Z}z\,n_{Z,z}.
\end{equation}
By multiplying both sides of Eq.(\ref{ionfrac}) by $1/n_{Z}$ ($n_{Z}$ is the number density (cm$^{-3}$) of the species of atomic number $Z$)  and factorizing $n_{e}$, the system of equations simplifies to
\begin{equation}
\label{ionfrac2}
S_{Z,z-1}x_{Z,z-1}-(S_{Z,z}+ \alpha_{Z,z})x_{Z,z}+\alpha_{Z,z+1}x_{Z,z+1}=0
,\end{equation}
where $x_{Z,z}=n_{Z,z}/n_{Z}$ is the ion fraction (which varies between 0 and 1). This system of equations may 
be cast into the matrix form
\begin{equation}
\label{matrix}
\mbox{AX}=0,
\end{equation}
where $\mbox{X}$ is a vector comprising all ion fractions $x_{Z,z}$ and $\mbox{A}$ is a tridiagonal matrix with 
elements $S_{Z,z-1}$, $-(S_{Z,z}+ \alpha_{Z,z})$, and $\alpha_{Z,z+1}$ at each row populating the diagonal band. 
The solution of this system of equations is straightforward using any Gauss-elimination method with or without 
pivoting. The final solution is then transformed into the ion density through
\begin{equation}
\label{iondensity}
n_{Z,z}=x_{Z,z}\,n_{Z}=x_{Z,z}\,A(Z)n_{H},
\end{equation} 
where $A(Z)=n_{Z}/n_{H}$ is the abundance of the species and $n_{H}$ is the hydrogen number density (cm$^{-3}$). 

The ionization rates, $S_{Z,z}$ , are determined by convolving $\sigma(E) v$ with the MB electron distribution function, 
\begin{equation}
f(E)dE=\frac{2E^{1/2}}{\pi^{1/2}(k_{B}T)^{3/2}} e^{-E/k_{B}T}dE,\end{equation}
and are given by
\begin{equation}
\label{rate_eii}
\langle \sigma v\rangle =\left(\frac{2}{m_{e}}\right)^{1/2}\int_{\Phi_{Z,z}}^{+\infty} \sigma(E) E^{1/2} f(E) 
dE\mbox{~~cm$^{3}$ s$^{-1}$},
\end{equation}
where $m_{e}$ is the electron mass (g), $\Phi_{z,z}$ is the ionization threshold (eV), and  $\sigma(E)$ is the 
electron-impact ionization cross section (cm$^{2}$).

Radiative and dielectronic recombination rates used in these calculations are based on calculations with the 
AUTOSTRUCTURE  code\footnote{amdpp.phys.strath.ac.uk/tamoc/DATA/} \citep{badnell2011} and are taken from 
\citet{badnell2006-rr} for \hii and \cii through \cvii recombining to \hi and \ci through \cvi, respectively. The 
radiative recombination rates have the functional
\begin{equation}
\alpha_{Z,z}^{\rm RR}=A\left[\left(\frac{T}{T_{0}}\right)^{1/2}\left(1+\left(\frac{T}{T_{0}}\right)^{1/2}\right)^{1-b}
\left(1+\left(\frac{T}{T_{1}}\right)^{1/2}\right)^{1+b}\right]^{-1}
,\end{equation}
where $A$ (cm$^{3}$ s$^{-1}$), $T_{0,1}$ (K), and $b$ (dimensionless) are fit coefficients. The latter is replaced 
by $b+C\exp(T_{2}/T)$ (C is dimensionless and $T_{2}$ is given in K) for low-ionization stages. The dielectronic 
recombination rates are given by the \citet{burgess1965} general formula
\begin{equation}
\alpha_{DR}^{\rm MB}=\frac{1}{(k_{B}T)^{3/2}}\sum_{j}c_{j} e^{-E_{j}/(k_{B}T)} \mbox{~~cm$^{3}$ s$^{-1}$}
,\end{equation}
whose coefficients are taken from \citet{badnell2006-dr-h} for \hii and \cvi, \citet{bautista2007-dr-he} for \cv, 
\citet{colgan2004-dr-li,colgan2003-dr-be} for \civ and \ciii, respectively, and \citet{altun2004-dr-b} for \cii. 
\begin{figure*}[!ht]
        \centering
        \includegraphics[width=0.8\hsize,angle=0]{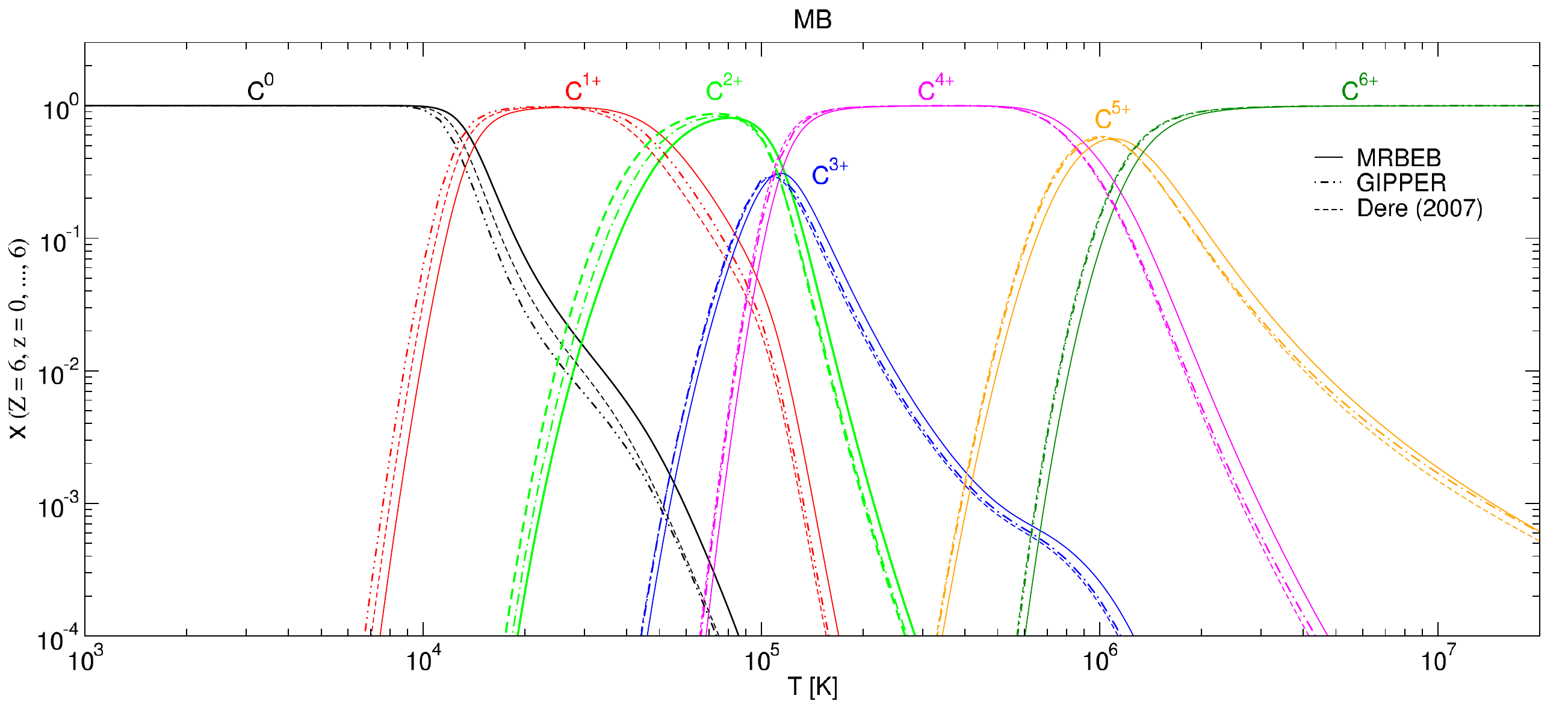}
        \caption{Carbon ionic fraction variation with temperature evolving under collisional ionization equilibrium 
        calculated with the MRBEB, GIPPER, and D2007 cross sections.}
        \label{carbon_ionfracs}
\end{figure*}

\begin{figure}[thbp]
        \centering
        \includegraphics[width=\hsize,angle=0]{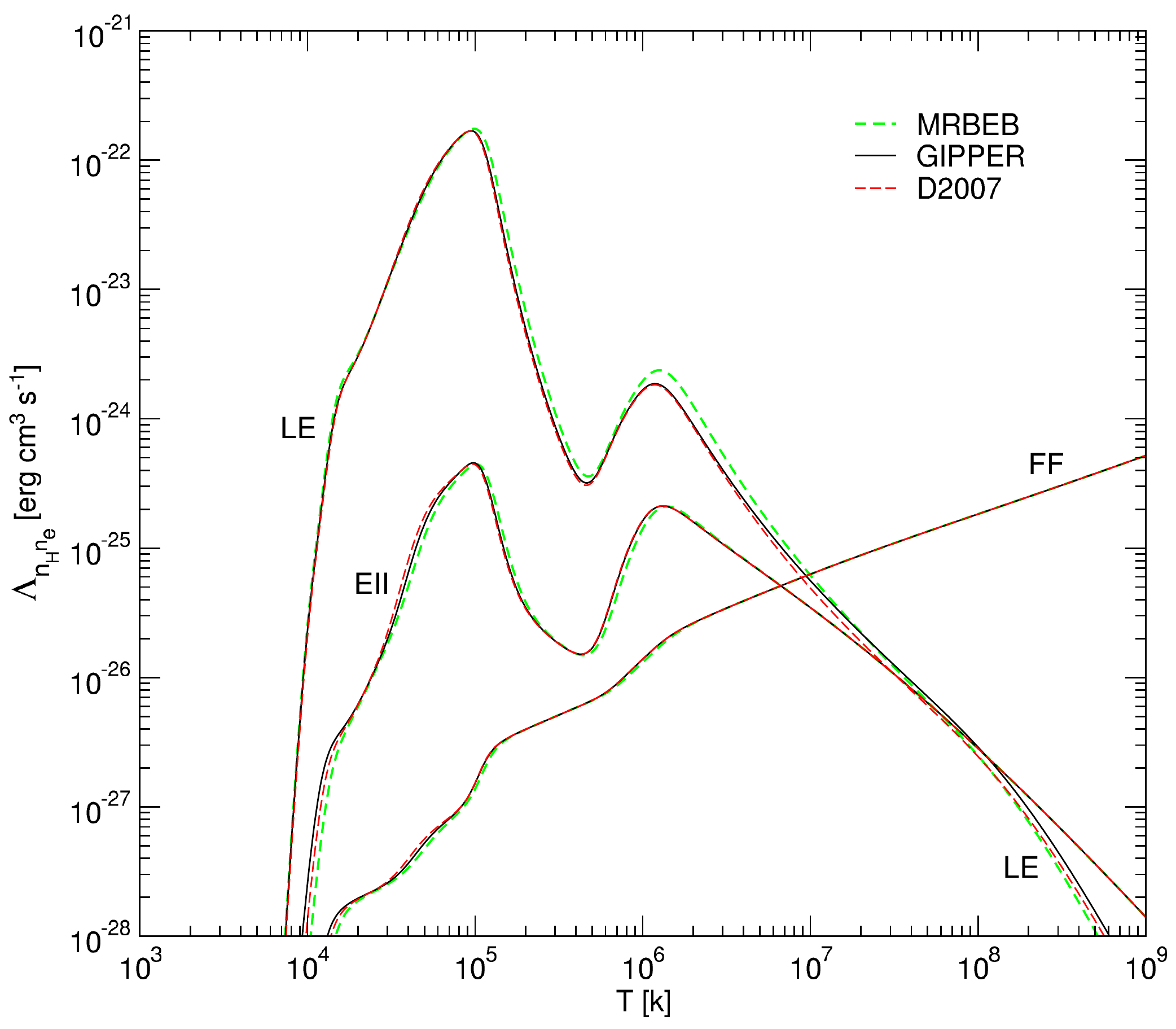}
        \caption{Comparison of the cooling (erg cm${3}$ s$^{-1}$), normalized to $n_{e}n_{H}$, due to electron-impact 
                ionization (EII), bremsstrahlung (FF), and line emission (LE)) in a gas parcel evolving under collisional 
                ionization equilibrium and calculated with MRBEB, GIPPER, and D2007 electron-impact ionization cross 
                sections.}
        \label{cooling_comparison}
\end{figure}

The cooling due to electron-impact ionization ($\Lambda^{\rm EII}_{Z,z}$), bremsstrahlung ($\Lambda^{\rm 
FF}_{Z,z}$), and line (permitted, forbidden, and semi-forbidden) emission ($\Lambda^{\rm LE}_{Z,z}$) shown in 
Fig.~\ref{cooling_comparison} are given by
\begin{equation}
\label{cool_eii}
\Lambda^{\rm EII}_{Z,z}=n_{e}n_{H}A(Z)x_{Z,z}S_{Z,z} \Phi_{Z,z}\mbox{~~~~erg cm$^{-3}$ s$^{-1}$},
\end{equation}
with $S_{Z,z}$ denoting the ionization rate (cm$^{3}$\,s$^{-1}$) and $\Phi_{Z,z}$ is the ionization threshold (erg) 
of the ionizing ion, 
\begin{equation}
\label{cool_ff}
\Lambda^{\rm FF}_{Z,z}=C\, z^{2} n_{e}n_{H}A(Z)x_{Z,z}T^{1/2} \langle g_{_{\rm ff}}(\gamma^{2})\rangle 
\mbox{~~~~erg cm$^{-3}$ s$^{-1}$},
\end{equation}
where $C=1.4256\times 10^{-27}$ erg cm$^{3}$ s$^{-1}$ K$^{-1/2}$, $T$ is the temperature (K), 
$\gamma^{2}=z^{2}\mbox{Ry}/k_{B}T$ (Ry is the Rydberg energy) is the normalized temperature, and  $\langle 
g_{_{\rm ff}}(\gamma^{2})\rangle$ is the total free-free Gaunt factor \citep[see the details of its calculation in, 
e.g.,][]{avillez2015}, and
\begin{equation}
\label{cool_lines}
\Lambda^{\rm LE}_{Z,z}=n_{e}n_{H}A(Z)x_{Z,z}\sum_{i,j,i<j}n_{Z,z,j}A_{ji}E_{ij} \mbox{~~~~erg cm$^{-3}$  
s$^{-1}$},
\end{equation}
where $n_{Z,z,j}$ is the population density of level $j$ of the ion, $A_{ji}$ is the spontaneous decay rate from 
level $j$ to level $i$ ($i<j$), and $E_{ij}$ is the excitation energy between the two levels.

The level populations were calculated as described in \citet{avillez2018} by assuming that there is an equilibrium 
between excitation by electron impact and de-excitation by electron impact and spontaneous decay. Hence, the 
population of level $j$ is obtained from the equation
\begin{eqnarray}
\label{excited_state}
\sum_{m<j}C_{mj}^{e}n_{_{e}}n_{Z,z,m}+\sum_{n>j}\left(A_{nj}+C_{nj}^{d}n_{e}\right)n_{Z,z,n} &-& \nonumber \\
n_{Z,z,j}\left[\sum_{j<n}C_{jn}^{e}n_{e}+\sum_{j>m}\left(A_{jm}+C_{jm}^{d}n_{e}\right)\right] & = & 0,
\end{eqnarray}
coupled to the equation of mass conservation
\begin{equation}
\sum_{j}n_{Z,z,j}=n_{Z,z}.
\end{equation}
In these equations $C^{e}_{mj}$ and $C^{e}_{jn}$ denote the excitation rates (cm$^{3}$ s$^{-1}$) from levels 
$m$ to $j$ ($m<j$) and $j$ to $n$ ($j<n$), respectively, while $C^{d}_{nj}$ and $C^{d}_{jm}$ denote the 
de-excitation rates (cm$^{3}$ s$^{-1}$) from levels $n$ to $j$ and $j$ to $m$, respectively; $A_{nj}$ and 
$A_{jm}$ are the Einstein spontaneous decay coefficients (s$^{-1}$) from levels $n$ to $j$ and $j$ to $m$, 
respectively; and $n_{Z,z,m}$ and $n_{Z,z,n}$ are the population densities (cm$^{-3}$) of levels $m$ and $n$, 
respectively. The excitation and de-excitation rates are given by
\begin{equation}
C^{e}_{ij}=8.629\times 10^{-6}\,T^{-1/2}\,\omega^{-1}_{i}\, e^{-y}\, \Upsilon_{ij}(T)
\end{equation}
and
\begin{equation}
C^{d}_{ji}=8.629\times 10^{-6}\,T^{-1/2}\,\omega^{-1}_{j}\, \rotatebox[origin=c]{180}{$\Upsilon$}_{ji}(T),
\end{equation}
where $y=E_{ij}/k_{B}T$, $U=E_{e}/E_{ij}$ is the reduced electron energy ($E_{e}$ is the energy of the impacting 
electron), $\omega_{i}$ and $\omega_{j}$ are the statistical weights of levels $i$ and $j$, respectively, and $T$ 
is the temperature (K). The forms of the effective collision strength, $\Upsilon_{ij}(T)$, and of 
$\rotatebox[origin=c]{180}{$\Upsilon$}_{ji}(T)$ relate to the collision strengths and are given by
\begin{eqnarray}
\label{x1}
\Upsilon_{ij}(T) &= &y e^{y}\int_{1}^{+\infty}\Omega_{ij}(U)\, e^{-yU}dU\\
\rotatebox[origin=c]{180}{$\Upsilon$}_{ji}(T) &=& y \int_{0}^{+\infty}\Omega_{ji}(U^{\prime})\, 
e^{-yU^{\prime}}dU^{\prime}.
\end{eqnarray}

The wavelengths, coefficients of spontaneous transitions, and the effective collision strengths were taken from 
version 8.0.7 of the CHIANTI atomic 
database\footnote{\href{http://www.chiantidatabase.org}{http://www.chiantidatabase.org}}.

\subsection{Calculations and methods}

The evolution of a gas parcel is calculated as follows: (i) at each temperature calculate first the ionization and 
recombination rates and solve the system of equations (\ref{ionfrac2}) to obtain the ion fractions, (ii) determine 
the ions and the electron densities from Eqs. (\ref{iondensity}) and (\ref{electron}), respectively, and (iii) obtain the 
emissivity due to the different processes using Eqs. (\ref{cool_eii})-(\ref{cool_lines}). 

The numerical methods used in these calculations are the same as in \citet{avillez2018}, that is, (i) numerical 
integrations in semi-finite intervals, such as the ionization rates, were calculated with a precision of 
$10^{-15}$ using the double-exponential transformation method of \citet{takahasi1974,mori2001}, and (ii) the 
system of equations (\ref{ionfrac2}) is solved using a Gauss-elimination method with scaled partial pivoting 
\citep{cheney2008} and a tolerance of $10^{-15}$. 

\subsection{Results}

The carbon-ion fraction variations with temperature  calculated with the MRBEB, GIPPER, and D2007 ionization 
cross sections are displayed in Fig.~\ref{carbon_ionfracs}. Although there are some differences between the 
location of the ion profiles, the ionic structure shows similar properties in the three models: the same profiles for 
all the ions that are characterized with the dominance of \cvii above $10^{6}$ K, the dominance of \ci below $10^{4}$ K, 
and the classic \cv plateau resulting from the K-shell ionization potential. The differences between the three models  
are reflected in the small deviation to the right for the MRBEB model with regard to the others, while the 
GIPPER model shows profiles that in some cases are to the left of the D2007 profiles and in other cases to the 
right, but always to the left of the MRBEB profiles. The reason is that the threshold ionization energies in the 
MRBEB model are higher than those of the GIPPER and D2007 models, which leads to a delayed ionization of the 
carbon ions in comparison to the other cases.

Fig.~\ref{cooling_comparison} displays the cooling (erg cm$^{3}$ s$^{-1}$), normalized to $n_{e}n_{H}$, due to 
electron-impact ionization (EII), bremsstrahlung (FF) using the total Gaunt factors calculated in 
\citet{avillez2017}, for instance, and line (allowed, forbidden, and semi-forbidden) emission (LE) calculated with the three 
models (MRBEB, GIPPER, and D2007). Emissivities associated with different processes 
in different ranges in temperature overlap. For instance, above $10^5$ K, the bremsstrahlung in the three models 
overlaps, except around $10^6$. A complete overlap is visible above $2\times 10^6$ K in electron-impact ionization 
and bremsstrahlung. In the latter the overlap extends to $10^5$ K, except around $10^{6}$ K. Deviations 
among the three models are seen in electron-impact ionization below $10^{6}$ K and in line emission above 
$10^{5}$ K, where the second peak maximum noticeably increases. For lower temperatures the line 
emission seems to be the same for all the models. The excess in emissivity in the MRBEB model is similar to 
that seen in the ionization structure (i.e., the deviation to the right). Although there are deviations in the 
models, they predict in general the same behavior and profile for the emissivities due to the different processes.

\section{Discussion and final remarks}

We applied the modified relativistic binary encounter Bethe model to calculate the K- and L-shell 
ionization cross sections of the carbon atom and ions and compare their variation with energy of the impacting 
electron with those calculated using the GIPPER code and those published by \citet{dere2007}, which includes 
cross sections calculated with the flexible atomic code and the relativistic distorted wave approximation of 
\citet{fontes1999}, and experimental data. 

In general, the three sets of cross sections have a similar profile for \cii, \ciii, and \civ ions, and small deviations occur during the 
approach to the peak (\cii), in the peak (\ciii and \civ), and after the peak (\ciii and \civ) maximum. The total MRBEB cross sections 
for \cii - \cvi ions dominate the others for $T>10^{8}$ K because of the relativistic effects. The MRBEB and 
GIPPER cross sections have small deviations for \ci, \cii, \ciii, and \civ near and at the peak maximum, while the 
largest deviations occur in the approach to the peak maximum for \cv and \cvi .  The D2007 cross sections have 
the highest values for \cv and \cvi ions at peak maximum and at high energies up to $10^{8}$ K when the MRBEB 
cross sections take over. Although the FAC and GIPPER code use the distorted-wave method for the electron-impact ionization, the D2007 and GIPPER cross sections still show differences that can stem from the different 
numerical methods and the adopted atomic model.

We further explored the effects of the three sets of cross sections on the ion fractions of  the carbon ions and on 
the cooling due to electron-impact ionization, bremsstrahlung, and line emission by an optically thin plasma 
that evolves under collisional ionization equilibrium and cooling from a temperature of $10^{9}$ K. The three 
calculations, using the MRBEB, GIPPER, and D2007 cross sections, show deviations in the ion fractions of the 
same ion that decrease with increase in ionization degree: the strongest deviations occur in the lowest ionization states 
(\ci-\ciii), and the smallest deviations in the highest ionization states. These differences in the ion fractions propagate to 
the emissivities. At high temperatures the emissivities are similar in the three calculations, while in other 
temperature regimes noticeable differences are observed, for instance, in the second peak maximum of the line 
emission around $10^{6}$ K. The agreement between the emissivities calculated with 
the three sets of cross sections is nevertheless good overall.

The results show that a simple analytical model that only depends on one atomic parameter 
(the electron binding energy) is capable of providing electron-impact ionization cross sections similar to those 
calculated with the more sophisticated quantum mechanical methods in GIPPER and FAC. This shows that it is possible to build a 
database of cross sections associated with the atoms and ions of the ten most abundant elements in nature with the MRBEB 
method that can be used by any spectral emission code.

\begin{acknowledgements}
        This research was supported by the projects "Enabling Green E-science for the SKA Research Infrastructure 
        (ENGAGE SKA)" (M.A.; reference POCI-01-0145-FEDER-022217, funded by COMPETE 2020 and FCT) and 
        "Ultra-high-accuracy X-ray spectroscopy of transition metal oxides and rare earths" (M.G.; reference 
        PTDC/FIS-AQM/31969/2017, and SFRH/BPD/92455/2013 funded by FCT), and by the project 
        UID/FIS/04559/2013 (LIBPhys). Partial support for M.A. and D.B. was provided by the Deutsche 
        Forschungsgemeinschaft, DFG project ISM-SPP 1573. The calculations were carried out at the ISM - Xeon Phi 
        cluster of the Computational Astrophysics Group, University of \'Evora, acquired under project "Hybrid computing 
        using accelerators \& coprocessors - modelling nature with a novell approach" (M.A.), InAlentejo program, CCDRA, 
        Portugal. 
\end{acknowledgements}

\bibliography{bibliography}

\end{document}